\pdfoutput=1
\documentclass{sig-alternate}
\usepackage{multirow}
\usepackage{array}
\usepackage[monochrome]{color}
\usepackage{url}

\begin{document}
\title{Novel Power and Completion Time Models for Virtualized Environments}

\numberofauthors{1}
\author{
\alignauthor
Swetha P.T. Srinivasan and Umesh Bellur\\
    \affaddr{Dept. of Computer Science and Engineering}\\
    \affaddr{Indian Institute of Technology, Bombay}\\
    \email{\{swethapts,umesh\}@cse.iitb.ac.in}
}

\maketitle
\begin{abstract}
Power consumption costs takes upto half of operational expenses of datacenters making power management a critical concern. Advances in processor technology provide fine-grained control over operating frequency and voltage of processors and this control can be used to tradeoff power for performance. Although many power and performance models exist, they have a significant error margin while predicting the performance of memory or file-intensive tasks and HPC applications. Our investigations reveal that the prediction error is due in part to the fact that they do not take frequency AND CPU variations account, rather they just depend on the CPU by itself.

In this paper, we empirically derive power and completion time models using linear regression with CPU utilization and operating frequency as parameters. We validate our power model on several Intel and AMD processors by predicting within 2-7\% of measured power. We validate our completion time model using five kernels of NASA Parallel Benchmark suite and five CPU, memory and file-intensive benchmarks on four heterogeneous systems and predicting within 1-6\% of observed performance. We then show how these models can be employed to realize as much as 15\% savings in power while delivering 44\% better performance for applications deployed in a virtualized environment. 
\end{abstract}


\section{Introduction}
\label{ch:intro}
Power consumption still remains the greatest concern of data center administrators taking 30-50\% of the operational costs \cite{Qureshi2009}. Data centers today are equipped primarily with multicore machines which offer advanced power management techniques. Processor manufacturers such as Intel and AMD have introduced products such as AMD PowerNow!, AMD Cool'n'Quiet and Intel SpeedStep that incorporate Dynamic Voltage and Frequency Scaling (DVFS) and Dynamic Power Management such as clock gating. Intel Nehalam architecture included per-core Dynamic Frequency Scaling that enables each core with a separate Digital Phase-Locked Loop (DPLL) \cite{Kurd2009} for clock signal generation rather than using clock gating. Also, finer-grained processor frequency steps, operating points and sleep states have considerably reduced the idle power consumption and greatly expanded the dynamic power range of the processors to as much as 62\% \cite{swetechreport}. With the use of these hardware techniques along with OS-level configurations (CPUgovernors), the idle power consumed is as low as 38\% of the peak power \cite{swetechreport}, unlike the earlier architectures that drew about 70\% of the peak power \cite{Fan2007}. 
\begin{figure}[!htbp]
\centering
\includegraphics[scale=1]{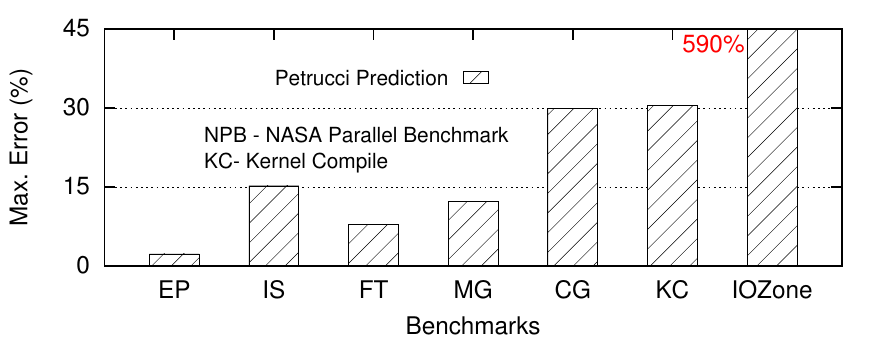}
\caption{Maximum error\% while predicting completion time for NASA Parallel benchmark suite and 2 other benchmarks using existing model}
\label{fig:ctErrorIntro}
\end{figure}

Almost all data centers today are virtualized and the tasks executing inside virtual machines (VMs) are isolated from each other. The hypervisor that facilitates the virtualization, also supports varying the frequencies of the processor cores the VMs are scheduled on as well as the CPU allocation of individual VMs. This gives us clear provisioning boundaries to work with and hence a good target to control power consumption of servers.


Power drawn by a system depends on the frequency at which the processor operates as well as the CPU utilization itself i.e., power varies by as much as 62\% across CPU usage and 45\% across the highest and the lowest frequencies \cite{swetechreport}. A thorough survey of existing literature on \emph{power models} raised two issues with respect to applicability on modern processors. 
\begin{itemize}
 \item Most models either focused on measuring the power consumed at the component-level using voltage and frequency supplied to the processor, or modeled power as a function of CPU and other resource utilizations but not both.
 \item Models that consider CPU and frequency as parameters fail to take into account the wide range of frequency settings offered and the kind of frequency scaling (clock gating vs DPLL).
\end{itemize}
Traditionally, the power consumed is inversely proportional to (and is thus traded off with) application performance which in a data center directly affects the SLAs guaranteed by a service provider. There is a vast collection of literature that quantifies the effect of either the CPU allocation or the processor frequency on the performance of benchmarks. The review of DVFS-aware \emph{performance models} for virtualized applications evoked following drawbacks. 
\begin{itemize}
 \item Most models either predicted performance of applications under frequency scaling or CPU reallocation that `simulated' frequency scaling. 
 \item Combined effect of CPU and frequency changes on the performance of applications are not analyzed thoroughly.
 \item Gaps in existing work are clearly evident from Figure \ref{fig:ctErrorIntro}. Prediction error for CPU-intensive tasks are low, about 30\% for memory-intensive applications and as high as 590\% for file-intensive benchmarks. 
 \item Need unified performance model for that considers both frequency and CPU allocation
\end{itemize}

Our ultimate goal is to provision VMs while satisfying the twin (and possibly competing) requirements of a power budget and VM performance. As an intermediate step, with a power and completion time model in place, we could optimize for one or the other of these requirements as well and use the resulting model to understand the effect that maximizing performance will have on power or minimizing power consumption will have on the performance. The two building blocks that allow us to predict power and performance are what we present in this paper. 

Our contributions towards power and completion time prediction in virtualized environments are the following.
\begin{itemize}
 \item Empirically establish that for a multicore processor, only the highest currently operating frequency affects the power. 
 \item Identify the minimal set of input parameters and empirically derived a model using those parameters for predicting power consumption of virtualized servers.
 \item Validate our derived model across heterogeneous processors with high accuracy.
 \item Propose and validate a completion time model that considers compute resource and frequency as parameters across different application types and heterogeneous systems with high precision. 
 \item Integrate our proposed models and demonstrate scenarios which lead to significant power savings and performance improvements while provisioning VMs.
\end{itemize}


The rest of the paper is organized as follows. Section \ref{sec:background} describes the background on the existing hardware-level power management techniques applicable to data centers. Existing power and performance models are presented in Section \ref{sec:RWork}. Section \ref{sec:WModel} provides the methodology, experimental setup and identifies the parameters needed for predicting the power of a virtualized server. Section \ref{sec:WModelDer} derives a power model with CPU\% and operating frequency as parameters and validates the model across heterogeneous servers. Section \ref{sec:PfModel} identifies compute resource and server frequency as the input parameters for predicting the performance of a task. Section \ref{sec:PfModelDer} derives a completion time model from existing work and validates the model across 4 systems and 10 benchmarks, including NASA Parallel Benchmark suite. Section \ref{sec:integration} presents two scenarios where the integration of our power and performance models leads to significant amount of power saved and performance speedup. Conclusions and future work are presented in Section \ref{sec:conclusions}.

\section{Background}
\label{sec:background}
The power consumption of a system is the rate at which the system performs work and has two components - $P^{total} = P^{idle} + P^{dynamic}$ (watts). $P^{idle}$ is the minimum power that is required by the system to remain active, irrespective of clock rate or usage. $P^{dynamic}$ is the power consumed while performing computations. $P^{dynamic}$ varies with clock rate, voltage supplied or utilization of the system. The dynamic power range of a system is defined as the ratio of the difference between peak power and idle power to the peak power. Energy consumption is total work done by system for a time duration i.e., $E = \Sigma ^t _{i=1}P(i)$ and is measured in watt-hours or joules. Data centers which use modern processors could achieve higher power conservation by understanding and effectively using the Hardware-assisted techniques currently offered. 
\subsection{Hardware-assisted Power Management}
Benini et al. \cite{Benini2000} classified power management of processors as (1) supply shutdown; (2) clock gating and (3) multiple and variable power supplies for individual components. Modern processors offer dynamic frequency scaling (DFS) either by clock gating or using separate clock signals for each core. DVFS is a powerful dynamic power management technique for conserving the processor power by reducing the voltage and the frequency depending on the CPU-resource utilization. The dynamic power of the processor is given by $P^{dynamic} \ \propto \ v^{2} f$ where $v$ is the supply voltage and $f$ is the frequency across the processor. A change in the voltage will have a greater impact on the power drawn than a change of the frequency. 

Single-core processors were initially designed with off-chip voltage regulators which caused tens of milliseconds delay during DVFS \cite{Zhao2011}. The voltage regulators were placed off the chip due to their bulky size and space restrictions of the processor. Global on-chip regulators were designed to moderate the voltage supplied to multicore processors. The voltage is set based on the core operating at the highest frequency. Frequency scaling is done by scaling the clock length, which enables the required threshold voltage to be set. Whereas, the frequencies of the remaining cores are scaled using clock gating i.e., stopping the cores for some cycles \cite{Yadav}. Commercial processors such as Intel Nehalem currently use a DPLL to scale the frequency of individual cores. However, the voltage is still set based on the highest frequency. 
CPUgovernor has been incorporated into the linux kernel to provide a variety of frequency profiles to the users i.e., the governor chooses which frequency to set based on the CPU utilization and a set of selection policies. The five profiles that are offered are - Conservative, Ondemand, Performance, Powersave and Userspace. Current version of CPUgovernors used by the hypervisors do not consider the performance achieved by the applications executing inside the VMs, the required performance i.e., SLA or design of processors. 

In this paper, we aim to understand the effect of DVFS technique on the power consumption of the machines and the performance attained by the tasks executing inside the VMs, in order to achieve higher power conservation without SLA violations for VM provisioning. 
\section{Related Work}
\label{sec:RWork}
Power Management has become an essential part of data center operations. Virtualization has aided in power conservation, as it enabled higher utilization using fewer number of servers. There is scope for further improvement. This section presents the existing power models for modern multicore processors and the available performance model of virtualized applications.
\subsection{Power Models}
Modeling the power consumption of a system is an essential phase in efficient power management. Earliest works in this area used the processor's power consumption as a proxy for modeling power drawn by the system. The CMOS circuits
that are used for building the processors derive dynamic power as $P^{dynamic} = aCV^{2}f$ where $a$ is the switching activity, $C$ is the capacitance, $V$ is the supply voltage and $f$ is the frequency or rate of clock signal of the processor. Fan et al. \cite{Fan02} established a linear relationship between power consumption and CPU utilization as 
$$P = C_{0} + C_{1}u$$where $u$ is the fraction of CPU utilization. Bellosa et al. \cite{Bellosa} like \cite{Isci2003}, \cite{Bertrana}, \cite{Singh2009}, used Performance Monitoring Counters (PMC) to model the power consumption of the processor. Their models focused on CPU utilization as processors consumed about 70\% of the total power supplied to the system. However, the models did not hold for non-CPU-intensive applications \cite{Rivoire2008}. Gurumurthi et al. \cite{Gurumurthi2002} designed SoftWatt simulator to correlate power consumption with utilization of four resources, namely - CPU, memory, I/O and disk. Though the simulation method aids in analyzing the power drawn by individual components, it is empirically infeasible as the modern day processors are made of millions of such components. 

Pedram and Hwang proposed WorkloadGen \cite{Pedram2011}, a workload generator to model the dynamic power management technique - DVFS. This model was based on two assumptions - (i) only two frequencies available - `high' and `low'. (ii) if a core is at the `low' frequency, that core is idle i.e., no process is scheduled on that core. However, there are commercial multicore processors that support a wide range of frequencies to choose from and the cores may be in use even at lower frequencies.

Petrucci et al. \cite{Petrucci2011} addressed this concern when they proposed a power model that support multiple frequency steps. The power $p_{ij}$ at any given utilization $u_{ij}$ is given as 
$$p_{ij}(u_{ij}) = Pm_{ij} + (PM_{ij} - Pm_{ij}) \cdot u_{ij}$$
$$Pm_{ij} = Pm_{i1} + (Pm_{iF_i} - Pm_{i1})\frac{(f_{ij} - f_{i1})^2}{(f_{iF_i} - f_{i1})^2}$$ 
$$PM_{ij} = PM_{i1} + (PM_{iF_i} - PM_{i1})\frac{(f_{ij} - f_{i1})^2}{(f_{iF_i} - f_{i1})^2}$$

where $Pm_{iF_i}$ and $Pm_{i1}$ are the idle power at maximum and minimum frequencies i.e., $f_{iF_i}$ and $f_{i1}$ respectively. $PM_{iF_i}$ and $PM_{i1}$ are the peak power at $f_{iF_i}$ and $f_{i1}$ respectively. It is to be noted that power is linearly proportional to the frequency or cubically proportional as voltage of the processor is set based on the frequency. Petrucci et al. \cite{Petrucci2011}, however assumed a \emph{quadratic} relationship between power and frequency. 

In this paper, we empirically establish power's linear dependency on the frequency and since our work is closely related to \cite{Petrucci2011}, we show how our model predicts power consumption more accurately than that given in \cite{Petrucci2011}. In the latest survey of power models \cite{Mobius}, Petrucci's model that we have used for comparison, is the only model cited which combines CPU\% and frequency. \textbf{As far as we know, no other paper have 
proposed a combination of these two parameters.} Consider the case of Intel i7 processor - it has a considerably low idle power - 38\% of the peak power. The remaining 62\% of the total power depends not only on the CPU utilization, but also the operating frequency. Operating at 100\% CPU at the lowest frequency consumes only 45\% of the power at the highest frequency. This highlights the need for modeling power based on the current CPU utilization \textbf{and} the operating frequency.
\subsection{DVFS-based Performance Models}
\label{perf} 
Hsu and Feng \cite{Hsu2005} empirically observed the effect of frequency change on the completion time of tasks by characterizing the compute-boundedness of each microbenchmark. They proposed a model that verified that the relative performance can be approximated to the relative number of instructions executed per second (MIPS) and the relative frequency. Dhiman et al. \cite{Dhiman2008} and Marinoni and Buttazzo \cite{Marinoni2007} experimentally verified that frequency changes have lesser effects on memory-intensive applications and minimally affect network- and disk-intensive applications. Wang and Wang \cite{Wang2011} used Model Predictive Control (MPC) theory to design the Controller that changes the CPU allocation of the VM and the frequency of the servers based on a power cap. Though their performance model considers both the CPU and frequency of the server as parameters, their experimentation were neither performed on non-CPU-intensive applications, which have lesser performance loss with the change in frequency, nor heterogeneous applications with varied SLAs. 

Non-CPU-intensive applications were again neglected by Petrucci et al. \cite{Petrucci2011} when they proposed a performance model that depends on the CPU utilization and frequency of the server. They assumed CPU as the bottleneck for \emph{httperf} tool and predicted the performance $r_{ij}$ for any given utilization $u_{ij}$ and frequency $f_{ij}$ using the following equation. 

\begin{center}$r_{ij} (u_{ij} ) = R_{iF_i} \cdot u_{ij} \cdot \frac{f_{ij}}{f_{iF_i}}$\end{center}

where $R_{iF_i}$ is the performance at maximum frequency $f_{iF_i}$ and CPU utilization. This model was used in Figure \ref{fig:ctPredictCDFmulti} to emphasize the gap in existing work. In this paper, we show how our model predicts completion time more accurately than that given in \cite{Petrucci2011}. 

Another flavor in literature is to use OS or hypervisor-driven techniques to provide frequency scaling. Nathuji and Schwan, in VirtualPower \cite{Nathuji2007}, proposed `Soft Scaling' - a technique where VM will execute at the required frequency using CPU scheduling policy rather than hypervisor changing the frequency of the processor. Many other authors such as Kamga et al. \cite{Kamga2011} and Wen et al. \cite{Wen2010} proportionally reallocated CPU to simulate frequency changes. 

We would like to highlight here that the literature either dealt with characterizing performance for frequency scaling or used CPU reallocation to `simulate' frequency scaling but \emph{not both}. This is the main drawback our proposed completion time model overcomes. \textbf{As far as we know, we are the first to provide a unified model that predicts the completion time where frequency and CPU allocation are independently reconfigurable.} Such a model is particularly useful for virtualized environments for provisioning VMs without performance violations.

\section{Modeling Power Consumption of Virtualized Servers}
\label{sec:WModel}
Modern processors offer a minimum of 2 (AMD x4 9550) and up to 10 (Intel i7 2600) frequency steps. As discussed earlier, there is a need to consider the effect of operating frequency and the CPU utilization on the dynamic power of the system. Our methodology for empirically modeling power drawn by virtualized servers is described below.
\begin{itemize}
 \item Observe the power consumption trend of a system and the effect of various parameters such as number of VMs, CPU utilization, frequency of the cores, etc.
 \item Identify the key parameters that contribute to the power consumption.
 \item Empirically derive the power model through regression of the input parameters.
 \item Validate the model on other systems which are heterogeneous in terms of processor architecture, processor manufacturers and chassis. 
\end{itemize}
This section details the experimental setup to collect data to derive the power model. The input parameters of the model are determined through empirical evidence and regression is used to derive the model for Intel i7 2600. 
\subsection{Experimental Setup}
\label{expW}
The experiments in this paper are performed on 5 virtualized systems listed in Table \ref{tab:sys}. All the systems operate on linux kernel 3.2.0-23-generic-pae and QEMU Kernel-based Virtual Machine 1.0 hypervisor. The VMs also operate on linux kernel 3.2.0-23-generic-pae. The power is measured using KryKard ALM 10 \cite{ALM10} connected to the input supply of the systems. We experiment with a pseudo microbenchmark performing long double multiplication operations. The four VMs are pinned to four different cores and each VM simultaneously executes the benchmark. The frequency of the cores are set using \texttt{cpufreqd 2.4.2} and the CPU allocation is capped using \texttt{cpulimit 1.1}. All experiments are carried out with the processor frequency and CPU allocation of VMs as independent variables. 
\begin{table}[!htbp]
\vspace{-0.5cm}
\caption{Configuration of systems under test}
\begin{center}
\begin{tabular}{|c|p{0.8cm}|p{0.8cm}|p{1cm}|c|}
\hline
\textbf{System}	&	$f_{min}$ (GHz)	&	$f_{max}$ (GHz)	&	$f$ step (MHz)	&	Other $f$s	\\	\hline
\textbf{i7 2600}	&	1.6	&	3.4	&	200	&		\\	\hline
\textbf{i5 760}	&	1.86	&	2.79	&	133	&	1.2, 2.79+GHz	\\	\hline
\textbf{Xeon E5507}	&	1.6	&	2.26	&	133	&		\\	\hline
\textbf{Xeon E5520}	&	1.6	&	2.26	&	133	&	2.26+GHz	\\	\hline
\textbf{AMD 9550}	&	1.1	&	2.2	&	1100	&		\\	\hline
\end{tabular}
\label{tab:sys}
\end{center}
\vspace{-0.5cm}
\end{table}
\subsection{Number of VMs as a Parameter}
\begin{figure}[!htbp]
\vspace{-0.5cm}
\begin{center}
 \includegraphics[scale=1]{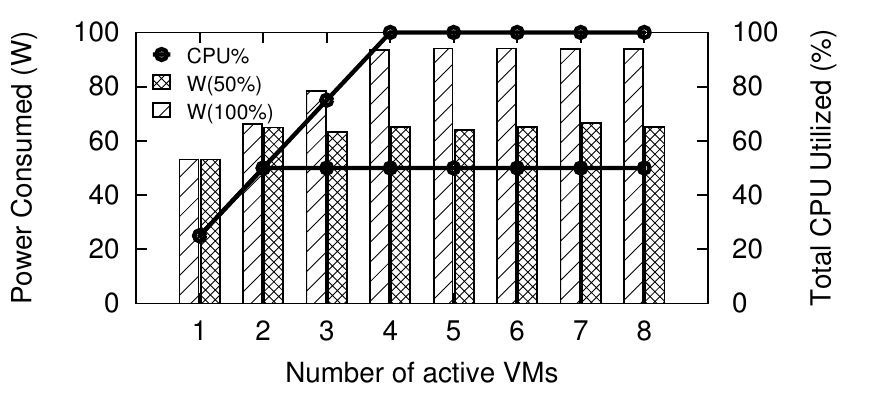}
\vspace{-0.5cm}
\caption{Power, CPU \% vs \# VMs on Intel i7}
\label{fig:numVM}
\vspace{-0.5cm}
\end{center}

\end{figure}

In this subsection, we look at whether the number of active VMs is a parameter that affects the power. The correlation between the number of active VMs and the power is determined by increasing the VMs running long double multiplication operations and the power consumption is observed for Intel i7 2600. In the first scenario, the VMs as allocated a cap of 50\% of the total CPU and in the second scenario, no such cap is imposed. The power remains a constant after 2 VMs for 50\% overall CPU cap of the system and 4 VMs for 100\% i.e., no CPU cap on the quad core processor, as shown in Figure \ref{fig:numVM}. This suggests that rather than number of active VMs, the CPU utilization of the system is a more accurate parameter to predict the power consumed by the system. 
\subsection{Individual Core Frequencies as a Parameter}
We measure the power consumption of a system on which 4 VMs running long double multiplication operations are executing. Each of the 4 VMs are pinned to one each of the quad cores of Intel i7 and are allocated 100\% of the core. Let $f_n$ be the frequency of core $c_n$. We operate each core on a combination of the following 4 frequencies - 3.4GHz, 2.6GHz, 2GHz and 1.6GHz. For example, configuration C2 has the first core $c_1$ operating at 3.4GHz, $c_2$ at 2GHz, $c_3$ at 1.6GHz and $c_4$ at 2.6GHz. Table \ref{corefi7} lists a few frequency configurations across the 4 cores and shows that even if one of the cores is operating at the highest frequency, the power consumption is the same (99 W). The lowest power of 51 W is reached only when all the cores are operating at 1.6GHz. Moreover, all the cores are homogeneous and the order of the cores does not affect the power i.e., power consumed when operating at C2 is the same as operating at C3 (99 W). Thus, it is \emph{sufficient to use the highest operating frequency as the input to model the power consumption} of the system rather than the individual operating frequencies of each core. \textbf{As far as we know, we are the first to empirically establish that the power is affected by only the highest operating frequency.}
\begin{table}[!htbp]
\vspace{-0.5cm}
\caption{Power consumption for different frequency settings}
\begin{center}
\begin{tabular}{|c||c|c|c|c||c|}
\hline
\multirow{2}{*}{\textbf{Config. No.}}	&	\textbf{$f_1$}	&	\textbf{$f_2$}	&	\textbf{$f_3$}	&	\textbf{$f_4$}	&	\textbf{Power}\\ \cline{2-6}
& \multicolumn{4}{c||}{\textbf{(GHz)}} & \textbf{(W)} \\ \hline
C1	&	3.4	&	3.4	&	3.4	&	3.4	&	99	\\ \hline
C2	&	3.4	&	2.0	&	1.6	&	2.6	&	99	\\ \hline
C3	&	1.6	&	2.6	&	3.4	&	2.0	&	99	\\ \hline
C4	&	1.6	&	3.4	&	1.6	&	1.6	&	99	\\ \hline
C5	&	2.6	&	1.6	&	2.0	&	2.0	&	73	\\ \hline
C6	&	1.6	&	1.6	&	2.6	&	1.6	&	73	\\ \hline
C7	&	1.6	&	2.0	&	2.0	&	2.0	&	59	\\ \hline
C8	&	1.6	&	2.0	&	1.6	&	1.6	&	59	\\ \hline
C9	&	1.6	&	1.6	&	1.6	&	1.6	&	51	\\ \hline
\end{tabular}
\end{center}
\vspace{-0.5cm}
\label{corefi7}
\end{table}
%
\subsection{Selection of Input Parameters}
It is evident from the above experiments that in order to predict the power consumption of a virtualized server, the CPU utilization and the highest currently operating frequency are required. Moreover, voltage of the processor is not considered as modifying the manufacturer-set voltage settings would reduce the life time of the CPU and void the processor warranty. We also assume that the VMs are pinned evenly across all the cores. Therefore, the CPU\% and highest currently operating frequency are the two input parameters of our power model. We now proceed to derive our power model in the next section. 
\section{Derivation of Power Model for Virtualized Server}
\label{sec:WModelDer}
Our aim is to build a model to predict the power consumption of a server for a given CPU utilization and frequency settings with minimal inputs required. Let $P^{cpu}_f$ be the power drawn at $cpu$ allocation and $f$ operating frequency. The total power (TPower) of the system depends both on CPU and $f$, the basic power model is given as 
\begin{equation}
 TPower^{cpu}_{f} = P^{idle} + P^{dynamic}
 \label{eq:begin}
\end{equation}
\begin{figure}[!htbp]
\begin{center}
 \includegraphics[scale=1]{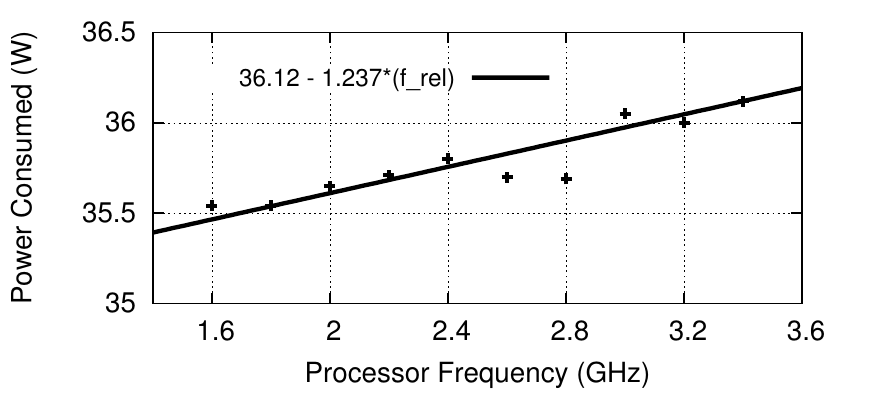}
\caption{Idle power consumption for Intel i7}
\label{fig:powerPloti7idle}
\end{center}
\vspace{-0.5cm}
\end{figure}
\begin{figure}[!htbp]
\begin{center}
\vspace{-0.5cm	}
 \includegraphics[scale=0.97]{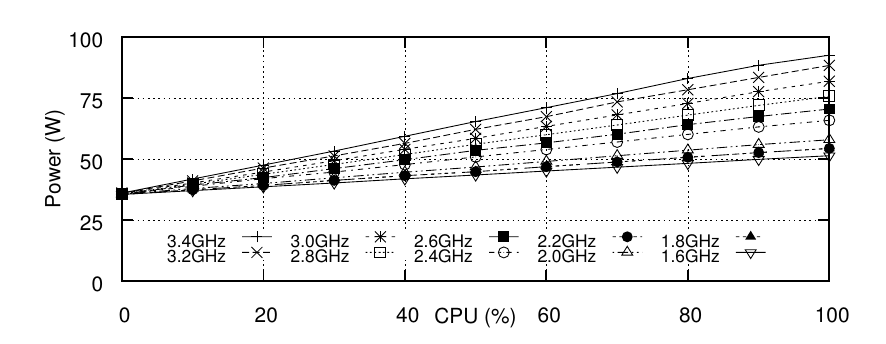}
\caption{Power trend of Intel i7 for various CPU and $f$ settings}
\label{fig:powerPloti7}
\end{center}
\vspace{-0.5cm}
\end{figure}
where $P^{idle}$ is the idle power and $P^{dynamic}$ is the dynamic power consumed. We now aim to model the idle power if Intel i7. The $P^{idle}$ for all the frequencies were measured and plotted in Figure \ref{fig:powerPloti7idle}. It is evident from Figure \ref{fig:powerPloti7idle} that $P^{idle}$ is dependent on the frequency of the processor. Moreover, it is trivial that CPU utilization is 0\% and hence, does not contribute to $P^{idle}$. In order to make the model system-independent, we use relative change in the frequency as the input parameter instead of the absolute frequency, and apply linear regression on the $P^{idle}$ values. With a linear fitting of 0.91 for the coefficient of determination, $R^2$, it is established that $P^{idle}$ is linearly dependent on the relative frequency and is modeled as \begin{equation}
P^{idle} = P^{0.0}_{f_{max}} - \alpha \cdot (\frac{f_{max} - f}{f_{max}})
\label{eq:idle}
\end{equation}
where $P^{0.0}_{f_{max}}$ is the idle power at $f_{max}$ and $\alpha$ = 1.237 (watts) is the idle power slope of Intel i7. Therefore, substituting Equation \ref{eq:idle} in Equation \ref{eq:begin}, we get \begin{equation}
\label{eq:alpha}
 TPower^{cpu}_{f} = P^{0.0}_{f_{max}} - \alpha \cdot (\frac{f_{max} - f}{f_{max}}) + P^{dynamic}
\end{equation}

We now empirically analyze and explain our intuition behind deriving a model for $P^{dynamic}$. Figure \ref{fig:powerPloti7} shows the power consumed for different CPU\% and frequency of the system. Intel i7 2600 offers 61.6\% of $P^{dynamic}$ and power varies by as much as 44.5\% across $f_{min}$ and $f_{max}$ at 100\% CPU utilization. It is clear that the power trend for each frequency is linearly proportional to the CPU\% i.e., $P^{dynamic}_f \propto cpu$ or $P^{dynamic}_f = \beta_f \cdot cpu$ for some $\beta_f$ (watts) at frequency $f$. Table \ref{fittingi7} shows the fitness of the above equation along with $\beta_f$ values for the ten frequency steps. The absolute difference between $P^{dynamic}_f$ and $\beta_f$ is very low for most frequencies. Therefore, Equation \ref{eq:alpha} can be rewritten as \begin{equation}
\label{eq:basicPower}
TPower^{cpu}_{f} = P^{0.0}_{f_{max}} - \alpha \cdot (\frac{f_{max} - f}{f_{max}}) + \beta_f \cdot cpu
\end{equation}
\begin{figure}[!htbp]
\begin{center}
 \includegraphics[scale=1]{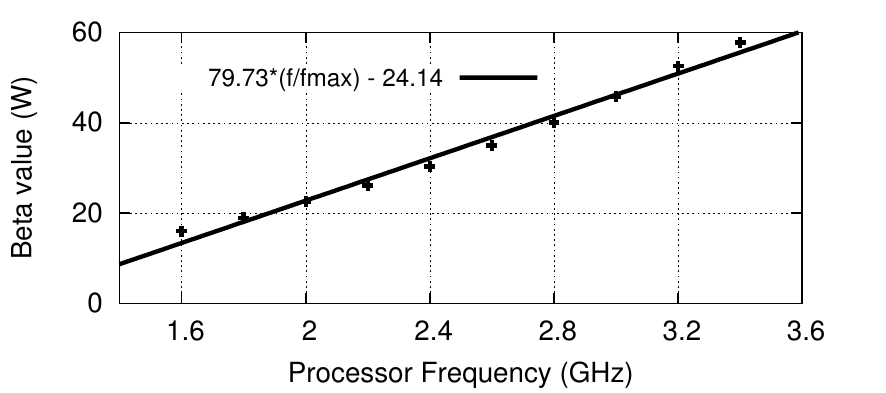}
\caption{$\beta_f$ values for Intel i7}
\label{fig:powerPloti7beta}
\end{center}
\vspace{-0.5cm}
\end{figure}

To use Equation \ref{eq:basicPower} as such for predicting power, ten $\beta_f$ values are required. Therefore, we further simplify our model by plotting $\beta_f$ values across different frequencies as shown in Figure \ref{fig:powerPloti7beta}. $\beta_f$ for any frequency is modeled as linearly dependent on the relative frequency with $R^2$ = 0.986 as 
\begin{table}[!htbp]
\caption{Linear Power Model fitting of Intel i7}
\begin{center}
\begin{tabular}{|c|c|c|c|c|}
\hline
\textbf{$f$ (GHz)} & \textbf{$\beta_f$} & \textbf{P$^{dyn}_f$} & $| \beta_f - P^{dyn}_f |$ & \textbf{$R^2$}\\ \hline
3.4 & 57.77 & 56.44 & 1.33 & 0.999\\ \hline
3.2 & 52.49 & 52.28 & 0.21 & 0.999\\ \hline
3.0 & 45.84 & 45.91 & 0.07 & 0.999\\ \hline
2.8 & 40.04 & 39.89 & 0.15 & 0.999\\ \hline
2.6 & 34.93 & 34.77 & 0.16 & 0.999\\ \hline
2.4 & 30.3 & 30.17 & 0.13 & 0.999\\ \hline
2.2 & 26.14 & 25.81 & 0.33 & 0.999\\ \hline
2.0 & 22.47 & 22.24 & 0.23 & 0.999\\ \hline
1.8 & 18.95 & 18.82 & 0.13 & 0.999\\ \hline
1.6 & 16.0 & 15.82 & 0.18 & 0.999\\ \hline
\end{tabular}
\end{center}
\label{fittingi7}
\vspace{-0.5cm}
\end{table}

 \begin{figure*}[t]
\centering
\includegraphics[scale=1]{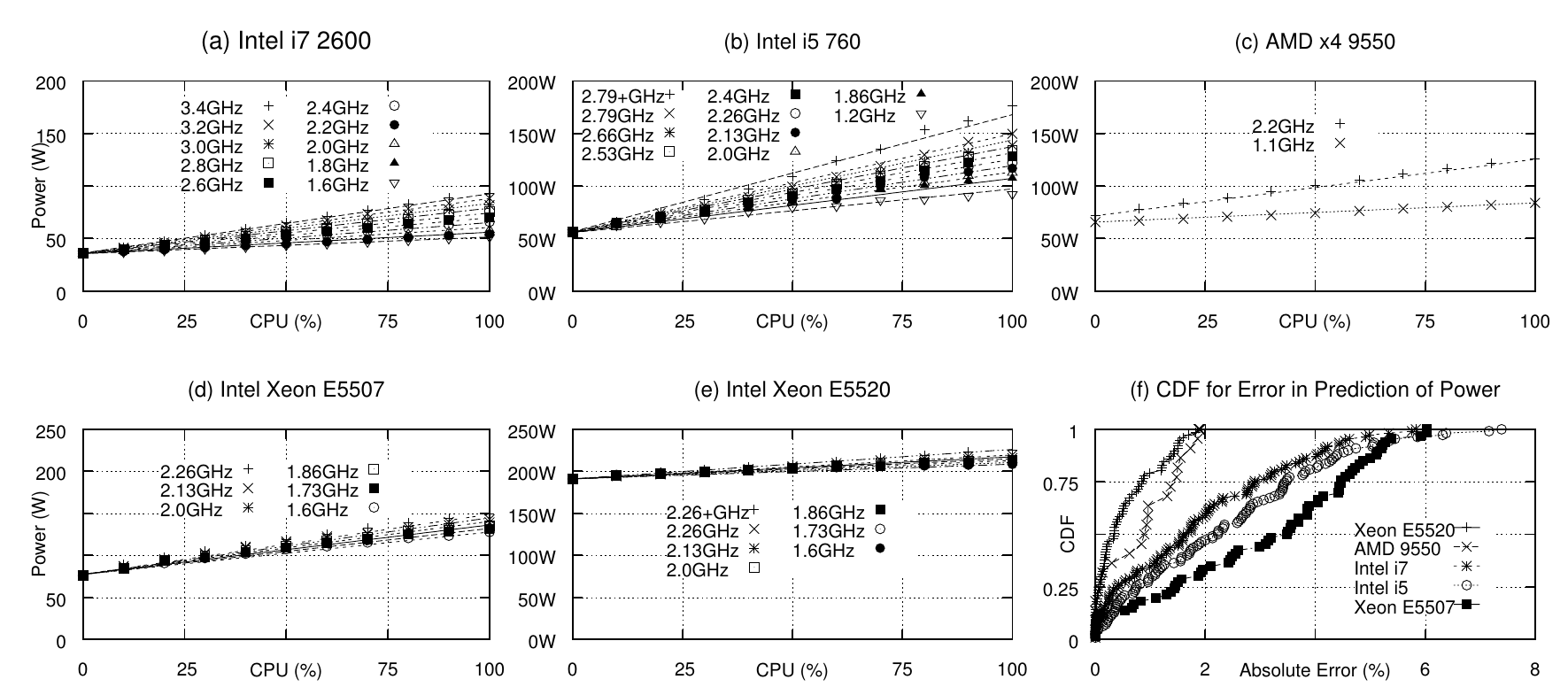}
\caption{Power Consumption trend of (a) Intel i7 2600 (b) Intel i5 760 (c) AMD x4 9550 (d) Dual Intel Xeon E5507 rack-mount (e) Dual Intel Xeon E5520 (f) CDF of error in prediction of power for 5 systems}
\label{fig:powerPlotmulti}
 \vspace{-0.5cm}
\end{figure*}

\begin{equation}
\label{eq:beta}
 \beta_f  = (A \frac{f}{f_{max}} + B)  
\end{equation}
where $A$ is the component of $P^{dynamic}$ that is dependent on the relative frequency and $B$ is not. $A$ and $B$ values are found to be 79.73 (watts) and -24.14 (watts), respectively. Also, A + B $\approx \ P^{dynamic}_{f_{max}}$. Using Equation \ref{eq:beta} in Equation \ref{eq:basicPower}, 
\begin{equation}
\label{eq:powerEq}
TPower^{cpu}_{f} = P^{0.0}_{f_{max}} - \alpha \cdot (\frac{f_{max} - f}{f_{max}}) + (A \cdot \frac{f}{f_{max}} + B)\cdot cpu
\end{equation}
The above equation requires only three calculated inputs - $A$, $B$ and $\alpha$ and the next subsection explains how they are obtained.  
\begin{table*}[t]	
\caption{Input values of Power Model for Intel and AMD systems and Prediction Errors}
\begin{center}
\begin{tabular}{|c|c|c|c|c|c|c|c|c||c|c||c|c|}
\hline
\textbf{Processor} 	& 	$P^{0.0}_{f_{min}}$  	&	$P^{0.0}_{f_{max}}$ 	&	$\alpha$ 	&	$P^{1.0}_{f_{min}}$ 	&	$P^{1.0}_{f_{max}}$ 	&	$A$	&	$B$	&	$P^{dyn}_{f_{max}}$ 	&	\multicolumn{2}{c||}{\textbf{\cite{Petrucci2011}'s Error\%}} 	&	\multicolumn{2}{c|}{\textbf{Our Error \%}} 	\\ 	\cline{2-13} 			
	&	\multicolumn{7}{c|}{(Watts)} 													&	($\%$) 	&	Avg 	&	Max 	&	Avg 	&	Max 	\\ \hline
 i7 2600	&	35.54	&	36.14	&	1.09	&	51.36	&	92.56	&	76.72	&	-20.28	&	61.60	&	3.83	&	9.68	&	1.89	&	5.83	\\ \hline
i5 760	&	55.75	&	56.28	&	1.59	&	107.25	&	149.72	&	125.82	&	-32.38	&	62.76	&	2.86	&	8.45	&	2.36	&	7.39	\\ \hline
 E5507	&	77.28	&	77.46	&	0.61	&	127.77	&	148.82	&	70.95	&	0.41	&	48.07	&	3.01	&	6.54	&	3.02	&	6.03	\\ \hline
 E5520	&	191.01	&	191.01	&	0.00	&	208.38	&	219.45	&	37.638	&	-9.198	&	12.95	&	2.41	&	5.89	&	0.97	&	2.00	\\ \hline
AMD 9550	&	65.63	&	71.62	&	11.98	&	83.88	&	125.44	&	71.14	&	-17.32	&	50.86	&	0.83	&	1.89	&	0.83	&	1.89	\\ \hline

\end{tabular}
\label{powerInput}
\end{center}
 \vspace{-0.5cm}
\end{table*}

\subsection{
Obtaining A, B and $\alpha$ values}
The $A$, $B$ and $\alpha$ values are calculated using Equations \ref{eq:A}, \ref{eq:B} and \ref{eq:alpha2}. 
\begin{equation}
\label{eq:A}
A = ((P^{1.0}_{f_{max}} - P^{0.0}_{f_{max}})-(P^{1.0}_{f_{min}} - P^{0.0}_{f_{min}}))\cdot\frac{f_{max}}{f_{max}-f_{min}}
\end{equation}
\begin{equation}
\label{eq:B}
B = (P^{1.0}_{f_{max}} - P^{0.0}_{f_{max}})-A
\end{equation}
\begin{equation}
\label{eq:alpha2}
\alpha = (P^{0.0}_{f_{max}}-P^{0.0}_{f_{min}}) \cdot \frac{f_{max}}{f_{max}-f_{min}}
\end{equation}
where $P^{0.0}_{f_{min}}$ is the power consumed when system is 0\% CPU (idle) and at $f_{min}$, $P^{1.0}_{f_{min}}$ is the power when system is at 100\% CPU and at $f_{min}$ and $P^{1.0}_{f_{max}}$ is the power at 100\% CPU and $f_{max}$.

We require only 6 inputs - $f_{min}$, $f_{max}$, $ P^{0.0}_{f_{min}}$, $P^{0.0}_{f_{max}}$, $P^{1.0}_{f_{min}}$ and $P^{1.0}_{f_{max}}$ to calculate $A$, $B$ and $\alpha$ and to predict the power at a given CPU\% and $f$.
Section \ref{valWM} describes the experimental setup for validating our power model and enlists the accuracy of prediction of our power model across heterogeneous systems.


 \subsection{Validation of the Power Model}
 \label{valWM}

The power model that we derived in Equation \ref{eq:powerEq} is validated by predicting power of 4 heterogeneous machines - quad core i5 760 Desktop, dual Xeon E5507 rack mount server, dual Xeon E5520 blade server and AMD x4 9550 quad core Desktop. The systems are heterogeneous in terms of processor architecture, number of processors, processor manufacturers and chassis and aids in establishing the validity of our power model across a variety of systems. 

We observe the idle and total power at $f_{min}$ and $f_{max}$. $A$ and $B$ values are calculated using Equations \ref{eq:A} and \ref{eq:B} respectively, and $\alpha$ is obtained using Equation \ref{eq:alpha2}. Figures \ref{fig:powerPlotmulti}(a)-(e) show the power consumption trend for the 5 systems, including Intel i7. Table \ref{powerInput} provides the $A$, $B$, and $\alpha$ values. The deviation of predicted power from the measured power is calculated using Equation \ref{eq:error}. Figure \ref{fig:powerPlotmulti}(f) shows the accuracy of our power model and the latter part of Table \ref{powerInput} compares prediction accuracy of Petrucci's model \cite{Petrucci2011} with our proposed model across 5 systems. 
\begin{equation}
\label{eq:error}
Error\% = \frac{|Measured - Predicted|}{Measured} \cdot 100\% 
\end{equation}
The maximum error in prediction ranged from 1.89\% for AMD x4 9550 to 7.39\% for Intel i5. The `Turbo' mode of Intel i5 and Xeon 5520 were assumed to be three frequency steps (one step is 133MHz) higher than the highest frequency 2.793GHz and 2.261GHz respectively i.e, the turbo mode of i5 and E5520 were assumed to be 3.192GHz and 2.66GHz respectively. And the `low' mode of Intel i5, 1.2GHz is assumed to be 5/3 frequency steps lesser than the lowest frequency of 1.862GHz i.e., the `low' mode's frequency is assumed to be 1.64GHz \cite{Li2008}. 

The following were made for the measured power across 5 systems. 
\begin{itemize}
 \item Idle power of blade and rack-mount servers are significantly higher than desktop machines due to their chassis housing more components, including dual processors and fans and in the case of Xeon 5520, integrated management module and ethernet switch. 
 \item Idle power of the Intel processors varies by only about 1W but AMD x4 9550 supports a reduction of 6W between its $f_{min}$ and $f_{max}$.
\item Even though Intel i7 and AMD 9550 have similar $P^{dyn}$ at $f_{max}$ - 56W and 54W respectively, the dynamic range varies by about 11\% due to the very low $P_{idle}$ of Intel i7 of 35W. 
 \item Low error of Xeon 5520 is due to the high idle power (191 W) and a narrow dynamic range of 12.95\% while Xeon 5507's idle power (77 W) is only 40\% of Xeon 5520's and has a wider dynamic range of 48.07\%.
 \end{itemize}
 The following are the observations of predicting power of the 5 systems using our model and Petrucci's model \cite{Petrucci2011}, as shown in Table \ref{powerInput}. \begin{itemize}
 \item 4 out of 5 systems have lesser or the same average prediction error \% for our model than Petrucci's model. The 5$^{th}$ system has a higher average error of 0.01\%. 
 \item Petrucci's model does not predict as accurately as our model, given the same input for nearly every system with the highest error being 9.68\%.  This is because they had modeled power based on the \emph{square} of the relative change in the frequency while we model it \emph{linearly}.
 \item With the maximum error of 7.39\%, our power model is able to predict the power for various CPU utilizations and frequency configurations across heterogeneous processors, processor manufacturers and chassis. 
\end{itemize}
Thus, we derived a model to predict the power consumption of a system using only 2 frequencies and 4 power values as inputs. We, then, validated with 5 systems which are heterogeneous in terms of processors, processor manufacturers and chassis. The power model built in a virtualized environment ensured the inclusion of power consumption of memory and the reusabilty of the model for power optimal VM provisioning. However, the derived model will hold for non-virtualized environments such as bare metal, HPC and Hadoop clusters. Their power values can be predicted by supplying the new power values of the HPC and Hadoop systems. As a part of future work, our model can be extended to multiple processors by considering the highest operating frequency of each processor which would be highly relevant for HPC environment utilizing multi- and many-core processors. Our power model can be used for predicting power as well as setting a power budget for VM placement strategies. 


\section{A Completion Time Model for Virtual Machines}
\label{sec:PfModel}
\begin{figure*}[t]
\vspace{-0.5cm}
\centering
\includegraphics[scale=1]{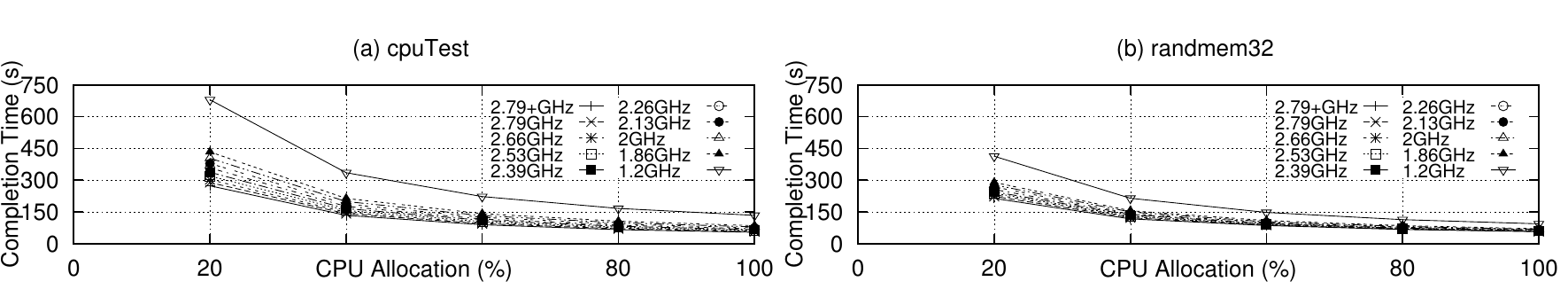}
\caption{Completion Time vs CPU\% and $f$ for Intel i5 (a) cpuTest (b) randmem32}
\label{fig:i5CTmulti}
\vspace{-0.5cm}
\end{figure*}
Service Level Agreements (SLAs) are the operative words of any data center service provider. Consider a scenario where the SLAs are based on the completion time of applications, which is a typical performance metric for HPC tasks and workloads that require batch processing such as Hadoop's Map-Reduce. The resources are allocated based on the required time for completion. In this section, we aim to understand the effect of compute-resource and frequency modifications on the execution time of the applications running inside VMs and experimentally derive a completion time model that can be used for provisioning VMs without violating execution time SLA constraints and reduce power consumption of the servers at the same time. Our methodology for empirically modeling completion time of VMs is described below.
\begin{itemize}
 \item Observe the completion time of VMs by varying the parameters that affect the completion time of benchmarks executing inside VMs
 \item Empirically derive the completion time model through regression of the input parameters.
 \item Validate the model with multiple benchmarks on other systems which are heterogeneous in terms of processor architecture, number of processors, processor manufacturers and chassis. 
\end{itemize}
In this section we identify the parameters that affect the completion time of a task executing inside a virtual machine and empirically derive a completion time i.e., the elapsed wall clock time, prediction model for Intel i5 760. In order to reduce the complexity of the model, the number of VMs is assumed to be the same as the number of cores of the server.
\subsection{Experimental Setup for Completion Time Model Derivation}
\label{expCT}
The experiments below were performed on virtualized Intel i5 760. For derivation of the completion time model, 4 VMs execute cpuTest for 11 billion iterative long double addition and multiplication and randmem32 \cite{randmem} benchmark that transfers data at increasing data sizes from and to caches and memory. The benchmarks are executed inside the VMs simultaneously by pinning them to four different cores. The average of the real value of \texttt{time} command of the 4 VMs is noted as the completion time for a specific CPU-frequency combination. The number of VMs is assumed to be 4 as it was established that with 4 VMs, peak CPU utilization and thus peak power is attained.

Figures \ref{fig:i5CTmulti} (a) and (b) graphically represent the completion time values of the 2 benchmarks. At 100\% CPU and the highest frequency, the completion time of cpuTest and randmem32 are almost the same - 57 and 58 seconds, respectively. At 20\% CPU and lowest frequency, the completion time of cpuTest and randmem32 vary drastically at 680 and 413 seconds. This substantial change in the completion time with respect to the frequency and CPU\% alterations are analyzed in Sections \ref{sec:freqCT} and \ref{sec:cpuCT}. A completion time model with CPU\% and frequency as parameters is derived and validated in Section \ref{sec:PfModelDer}.

\vspace{-0.1cm}

\subsection{Frequency of Processor as Parameter}
\label{sec:freqCT}
\begin{figure}
\centering
\includegraphics[scale=1]{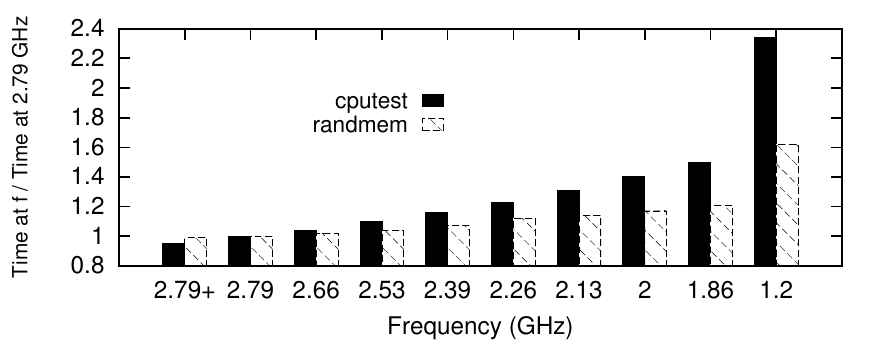}
\caption{Relative completion time vs $f$ for cpuTest, randmem32 on Intel i5}
\label{fig:i5cpurandmemrelCT}
\vspace{-0.5cm}
\end{figure}

The effect of frequency modification is observed for cpuTest and randmem32 at 100\% CPU on Intel i5 760 Desktop, as shown in Figure \ref{fig:i5cpurandmemrelCT}. The frequency change does not influence the relative completion time (CT) of randmem32 as much as cpuTest. This is due to cpuTest using the frequency-dependent hardware of the processor \cite{Venkatachalam2006}. 
Marinoni and Buttazzo \cite{Marinoni2007} have quantified the fraction of the frequency-dependency $U$ as 
$$ U = [ \frac{CT_{f_{min}}-CT_{f_{max}}}{CT_{f_{max}}} ] \cdot [ \frac{f_{min}}{f_{max}-f_{min}} ]$$
and the completion time for any frequency can be predicted using 
\begin{equation}
 \label{eq:freq}
 CT_f = [ U \cdot \frac{f_{max}}{f} + V ] \cdot CT_{f_{max}}
\end{equation}
where $U$ is the fraction of the application that is dependent on the frequency, $V$ is the fraction that is independent of the frequency changes and $U$ + $V$ =1. The $U$ value was empirically found to be 1.04 and 0.49 for cpuTest and randmem32 respectively. $U$ is architecture-dependent \cite{Marinoni2007} and needs to be recalculated for different processors. 

\subsection{CPU Allocation as Parameter}
\label{sec:cpuCT}

\begin{figure}[!htbp]
\vspace{-0.5cm}
\centering
\includegraphics[scale=0.96]{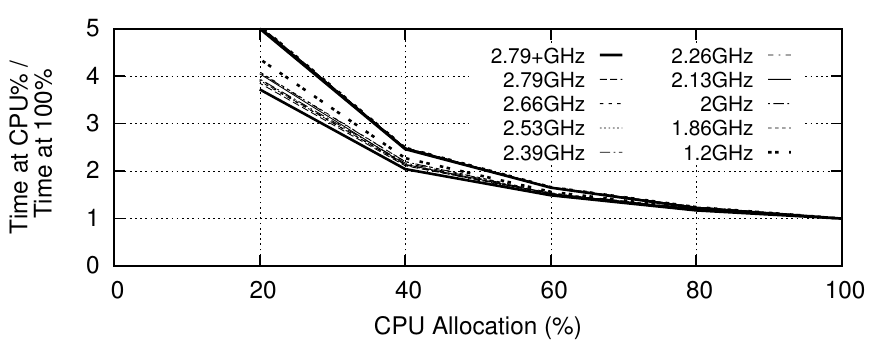}
\caption{Rel. completion time vs $cpu$ for cpuTest, randmem32 on Intel i5}
\label{fig:i5cpurandmemrelCTcpumono}
\end{figure}

Amdahl's law \cite{Amdahl1967}, one of the fundamental laws of Computer Architecture, is used to find the maximum expected improvement to an overall system when only part of the system is improved i.e., 
$$CT^{new} = [\frac{F_{enhanced}}{Speedup_{enhanced}} + (1-F_{enhanced})] \cdot CT^{old}$$
where $F_{enhanced}$ is the fraction of the application that is enhanced. Rewriting the above equation by expressing speedup $Speedup_{enhanced}$ as the CPU\% allocated, $$CT^{cpu} = [ \theta \cdot \frac{1.0}{cpu} + \mu ] \cdot CT^{1.0} $$ where $\theta$ quantifies the compute-boundedness of the application that is affected by the change in CPU allocation, $\mu$ is the CPU-independent part of the application and $\theta$ + $\mu$ = 1. For a CPU-intensive task, $\theta \approx $ 1. However, the above equation does not capture the combined effect of $f$ and CPU alteration on the completion time. While $\theta$ remains a constant for CPU-intensive tasks, it varies for non-CPU-intensive tasks with $f$ as shown in Figure \ref{fig:i5cpurandmemrelCTcpumono}. Therefore, $\theta$ has to be recalculated for every $f$.
\begin{equation}
 \label{eq:cpu}
 CT^{cpu}_{f} = [ \theta _{f}\cdot \frac{1.0}{cpu} + \mu _{f} ] \cdot CT^{1.0}_{f}
\end{equation}

In order to predict the completion time of tasks executing inside of VMs, the total CPU allocated and the frequency of the system are required and are thus, used as the input parameters of our completion time model. We now proceed to derive and validate our completion time model in the next section. 
\section{Derivation of Completion Time Model for Virtualized Environments}
\label{sec:PfModelDer}
Let $CT^{cpu}_f$ be the completion time for a given CPU utilization ratio $cpu$ and frequency $f$. Equation \ref{eq:freq} expresses the frequency dependency and Equation \ref{eq:cpu} gives the compute-boundedness of an application. Rewriting Equation \ref{eq:freq} at 100\% utilization as \begin{equation}
\label{eq:freq100}
CT^{1.0}_f = [ U \cdot \frac{f_{max}}{f} + V ] \cdot CT^{1.0}_{f_{max}}
\end{equation}
Equation \ref{eq:freq100} is plugged into Equation \ref{eq:cpu} to predict $CT^{cpu}_f$ as \begin{equation}
 \label{eq:timeEq}
 CT^{cpu}_{f} = [ \theta _{f}\cdot \frac{1.0}{cpu} + \mu _{f} ] \cdot [ U \cdot \frac{f_{max}}{f} + V ] \cdot CT^{1.0}_{f_{max}}
\end{equation}
 $U$ is calculated using $CT^{1.0}_{1.86GHz}$, $CT^{1.0}_{2.79GHz}$ and Equation \ref{eq:freq100} and is found to be 1.0 and 0.43 for cpuTest and randmem32 respectively. The $\theta_f$ and corresponding $R^2$ are given in Table \ref{ctcpu}. For the above equation to predict the execution time for any CPU\% and f, it requires ten $\theta_f$ values (for each $f$), in addition to $U$ and $CT^{1.0}_{f_{max}}$. Therefore, we aimed at reducing the inputs to the completion time model by applying linear regression to all the $\theta_f$ values of randmem32. 
\begin{table}[!htbp]	
\vspace{-0.5cm}
\caption{Verification for cpuTest and randmem32}
\begin{center}
\begin{tabular}{|c|c|c||c|c|}
\hline
$f$ (GHz) & \multicolumn{2}{c||}{\textbf{cpuTest U=1.0}} & \multicolumn{2}{c|}{\textbf{randmem32 U=0.43}} \\ \cline{2-5}
& $\theta_f$ & $R^2$ & $\theta_f$ & $R^2$ \\ \hline 
\textbf{2.79+}	&	1.0	&	0.999	&	0.71	&	0.999	\\	\hline
\textbf{2.79}	&	1.01	&	0.999	&	0.73	&	0.999	\\	\hline
\textbf{2.66}	&	1.01	&	0.999	&	0.74	&	0.999	\\	\hline
\textbf{2.53}	&	1.01	&	0.999	&	0.75	&	0.999	\\	\hline
\textbf{2.39}	&	1.01	&	0.999	&	0.75	&	0.999	\\	\hline
\textbf{2.26}	&	1.02	&	0.999	&	0.77	&	0.999	\\	\hline
\textbf{2.13}	&	1.01	&	0.999	&	0.78	&	0.999	\\	\hline
\textbf{2.0}	&	1.01	&	0.999	&	0.79	&	0.999	\\	\hline
\textbf{1.86}	&	1.01	&	0.999	&	0.79	&	0.999	\\	\hline
\textbf{1.2}	&	1.02	&	0.999	&	0.9	&	0.999	\\	\hline
\end{tabular}
\label{ctcpu}
\end{center}
\vspace{-0.5cm}
\end{table}
With $R^2$ as 0.978 suggests that $\theta_{f} = K\cdot( \frac{f_{max}}{f} - 1 ) + \theta_{f_{max}}$ and slope $K$ = 0.132 for randmem32. Since $\theta_{f} \approx \theta_{f_{max}}$ for all frequencies of cpuTest, $K$ is assumed to be 0. Moreover, $K$ can be approximated as $$K = (\theta_{f_{min}} - \theta_{f_{max}}) \cdot (\frac{f_{min}}{f_{max} - f_{min}})$$ and the calculated value was found to be $K$ = 0.12. Therefore, our model requires only $\theta_{f_{min}}$ and $\theta_{f_{max}}$ to calculate all other $\theta_f$ values. And how we calculate these values is explained in the next subsection. 

\begin{figure*}
\centering
\includegraphics[scale=1]{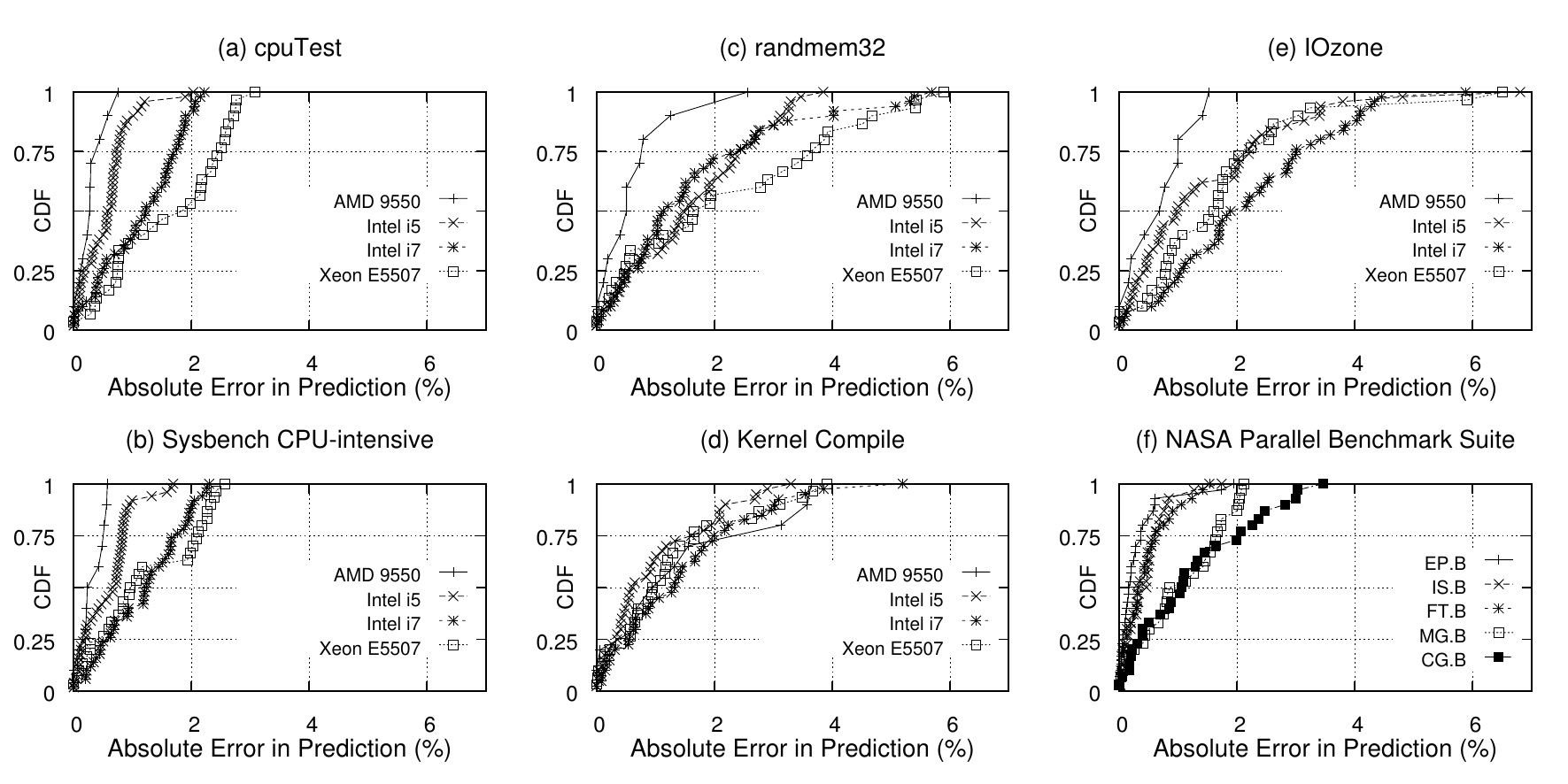}
\caption{CDF of error in predicting the completion time of (a) cpuTest (b) SysBench CPU (c) randmem32 (d) kernel compile (e) IOzone for AMD 9550, Intel i5, Intel i7 and Dual Xeon E5507 and (f) NASA Parallel Benchmark suite - Class B on Xeon E5507}
\label{fig:ctPredictCDFmulti}
\vspace{-0.5cm}
\end{figure*}

\subsection{Obtaining U and $\theta_f$ values}
U and $\theta_f$ values are calculated using Equations \ref{eq:U}, \ref{eq:thetafmax}, \ref{eq:thetafmin} and \ref{eq:thetaf}. 
\begin{equation}
\label{eq:U}
U = ( \frac{CT^{1.0}_{f_{min}} - CT^{1.0}_{f_{max}}}{CT^{1.0}_{f_{max}}} ) \cdot ( \frac{f_{min}}{f_{max}-f_{min}} )
\end{equation}
\begin{equation}
\label{eq:thetafmax}
 \theta_{f_{max}} = (\frac{0.x}{1.0 - 0.x}) \cdot (\frac{CT^{0.x}_{f_{max}} - CT^{1.0}_{f_{max}}}{CT^{1.0}_{f_{max}}})
\end{equation}
\begin{equation}
\label{eq:thetafmin}
 \theta_{f_{min}} = (\frac{0.x}{1.0 - 0.x}) \cdot (\frac{CT^{0.x}_{f_{min}} - CT^{1.0}_{f_{min}}}{CT^{1.0}_{f_{min}}})
\end{equation}
\begin{equation}
\label{eq:thetaf}
\theta_{f} =  (\theta_{f_{min}} - \theta_{f_{max}}) \cdot (\frac{f_{min}}{f_{max} - f_{min}}) \cdot( \frac{f_{max}}{f} - 1 ) + \theta_{f_{max}}
\end{equation}

We require only 7 inputs - $f_{min}$, $f_{max}$, $x\%$ CPU utilization, $ CT^{1.0}_{f_{min}}$, $CT^{0.x}_{f_{max}}$, $CT^{1.0}_{f_{max}}$ and $CT^{0.x}_{f_{max}}$ to predict the power at a given CPU\% and $f$. The following is the systematic procedure to calculate U for any arbitrary task/system. The same task is executed at 100\%CPU and 20\%CPU (say) on the minimum and maximum frequencies for a given application and system. VM cloning techniques such as SnowFlock \cite{SnowFlock} can be used to run cloned task on 4 different VMs having the 4 different configurations. The original VM is unaffected and continue to serve while the clones are executed using other configurations. Section \ref{expCTV} describes the experimental setup for validating our completion time model and Section \ref{valCT} enlists the accuracy of prediction of our completion time model and existing model used by Petrucci et al. \cite{Petrucci2011}, across heterogeneous applications and machines.
\subsection{Experimental Setup for Completion Time Model Validation}
\label{expCTV}

Experiments below were performed on virtualized Intel i5, Intel i7, Xeon E5507 and AMD 9550. We have experimented with 10  benchmarks - 5 kernels of NASA Parallel Benchmark \cite{nasa} - integer sort (IS), embarrassingly parallel (EP), conjugate gradient (CG), multi-grid (MG) and fast Fourier transforms (FT), all of class size B, and 5 microbenchmarks - SysBench CPU test \cite{sysbench} for finding the first 50,000 prime numbers, cpuTest \cite{cputest}, randmem32 \cite{randmem}, kernel compile 3.9.4 \cite{kc} and IOZone \cite{iozone} for creating and deleting 1000 files. These 10 benchmarks stress test CPU, memory, I/O and combinations of these resources. The benchmarks are executed inside VMs simultaneously by pinning them to individual cores. Class B of NPB suite was executed inside a single VM using all the cores of Xeon E5507 to empirically validate our completion time model on HPC benchmarks. Frequencies of the `low' and `turbo' modes of Intel i5 are assumed to be 1.197GHz and 2.926GHz respectively. 

\begin{table}[!htbp]	
\vspace{-0.5cm}
\caption{Validation of completion time model using 10 benchmarks on AMD 9550, Intel i5, i7 and E5507}
\begin{center}
\begin{tabular}{|c||c||c|c||c|c|}
\hline
\multirow{3}{*}{\textbf{Task}} 	&	\multirow{3}{*}{\textbf{System}} 	&	\multicolumn{4}{c|}{\textbf{ Error \% using}}			\\ \cline{3-6}				
	&		&			\multicolumn{2}{c||}{\textbf{Petrucci \cite{Petrucci2011}}}	& 	\multicolumn{2}{c|}{\textbf{Our Model}}	\\ \cline{3-6}		
	&		&	\textbf{Avg}	&	\textbf{Max}	&	\textbf{Avg}	&	\textbf{Max}	\\ \hline
\multirow{4}{*}{\textbf{cpuTest}}	&	 \textbf{AMD}	&	1.13	&	3.84	&	0.3	&	0.75	\\ \cline{2-6}
	&	 \textbf{i5}	&	0.55	&	2.00	&	0.57	&	2.03	\\ \cline{2-6}
	&	 \textbf{i7}	&	1.58	&	5.95	&	1.15	&	2.22	\\ \cline{2-6}
	&	 \textbf{E5507}	&	2.47	&	7.81	&	1.64	&	3.08	\\ \hline\hline
											
	&	 \textbf{AMD}	&	1.26	&	4.13	&	0.32	&	0.57	\\ \cline{2-6}
\multirow{1}{*}{\textbf{sys}}	&	\textbf{ i5}	&	0.52	&	2.16	&	0.59	&	1.69	\\ \cline{2-6}
\multirow{1}{*}{\textbf{bench}}	&	\textbf{ i7}	&	1.57	&	5.87	&	1.19	&	2.3	\\ \cline{2-6}
\multirow{1}{*}{\textbf{CPU}}	&	 \textbf{E5507}	&	1.26	&	8.19	&	0.32	&	2.56	\\ \hline\hline
											
	&	 \textbf{AMD}	&	19.32	&	37.63	&	0.70	&	2.57	\\ \cline{2-6}
\multirow{1}{*}{\textbf{rand}}	&	 \textbf{i5}	&	29.29	&	70.24	&	1.63	&	3.85	\\ \cline{2-6}
\multirow{2}{*}{\textbf{mem32}}	&	 \textbf{i7}	&	44.86	&	83.55	&	1.68	&	5.69	\\ \cline{2-6}
	&	 \textbf{E5507}	&	17.04	&	36.76	&	2.24	&	5.90	\\ \hline\hline
											
	&	 \textbf{AMD}	&	4.76	&	8.26	&	1.01	&	2.98	\\ \cline{2-6}
\multirow{1}{*}{\textbf{Kernel}}	&	 \textbf{i5}	&	4.89	&	14.38	&	1.01	&	3.30	\\ \cline{2-6}
\multirow{2}{*}{\textbf{Compile}}	&	 \textbf{i7}	&	11.04	&	17.66	&	1.47	&	5.20	\\ \cline{2-6}
	&	 \textbf{E5507}	&	17.55	&	30.37	&	1.29	&	3.91	\\ \hline\hline
											
\multirow{4}{*}{\textbf{IOzone}}	&	 \textbf{AMD}	&	231.46	&	854.99	&	0.71	&	1.52	\\ \cline{2-6}
	&	 \textbf{i5}	&	204.84	&	1047.79	&	1.47	&	6.81	\\ \cline{2-6}
	&	 \textbf{i7}	&	227.20	&	966.14	&	2.19	&	5.88	\\ \cline{2-6}
	&	 \textbf{E5507}	&	168.50	&	590.01	&	1.77	&	6.51	\\ \hline\hline
											
											
\multirow{1}{*}{\textbf{NPB.EP}}	&	\multirow{5}{*}{\textbf{E5507}}	&	0.55	&	2.23	&	0.32	&	1.94	\\ \cline{3-6}
\multirow{1}{*}{\textbf{NPB.IS}}	&		&	7.25	&	15.13	&	0.46	&	1.74	\\ \cline{3-6}
\multirow{1}{*}{\textbf{NPB.FT}}	&		&	4.23	&	7.86	&	0.46	&	1.53	\\ \cline{3-6}
\multirow{1}{*}{\textbf{NPB.MG}}	&		&	7.31	&	12.23	&	1.07	&	2.11	\\ \cline{3-6}
\multirow{1}{*}{\textbf{NPB.CG}}	&		&	13.15	&	29.89	&	1.28	&	3.46	\\ \hline

\end{tabular}
\label{ctval}
\end{center}
\end{table}

 \subsection{Validation of Completion Time Model} 
 \label{valCT}
Figures \ref{fig:ctPredictCDFmulti} (a)-(f) show the CDF of the error in predicting the completion time for 10 benchmarks on 4 systems using our completion time model. Table \ref{ctval} compares the prediction of our model  with  existing work. We observed the following while predicting the completion time. 
\begin{itemize}
 \item Dependency on the processor frequency i.e., the value of $U$ was $\approx$ 1 for both the CPU-intensive benchmarks and $\approx$ 0 for both the I/O-intensive benchmarks across the 4 heterogeneous servers.
 \item $U$ value vary from 0.25 to 0.52 for randmem32 and 0.45 to 0.84 for kernel compile. Therefore, $U$ has to be calculated for individual applications on each system. 
 \item Completion time monotonically increased from 100\% to 20\% for all frequencies, benchmarks and systems, except for kernel compile. While it still monotonically increased from 80\% to 20\%, compilation took longer at 100\% than 80\%. This led to a maximum error in prediction about 16\% on Intel i5, i7 and AMD 9550. This outlier could be attributed to the bottleneck created by CPU on the memory and file subsystems.
 \item To overcome the above exception, the base of the prediction to 80\%, i.e., the following equation was used to predict the time for kernel compilation where $U$ is obtained from $CT^{0.8}_{f_{max}}$ and $CT^{0.8}_{f_{min}}$
 $$ CT^{cpu}_{f} = [ \theta _{f}\cdot \frac{0.8}{cpu} + \mu _{f} ] \cdot [ U \frac{f_{max}}{f} + V ] \cdot CT^{0.8}_{f_{max}}$$
 \item The above method for prediction led to a maximum error of 5.2\% across of 20\% to 80\% CPU but as much as 39\% for 100\% CPU. This outlier could also be attributed to bottlenecks created on memory and file subsystems.
 \item Our average error is consistently lesser than \cite{Petrucci2011} for compute as well as other-resource-intensive tasks. 
 \item Compared to our model, \cite{Petrucci2011} predicted one \emph{magnitude} worse than measured values for memory-intensive and more than \textbf{two magnitudes} worse for file-intensive benchmarks.
 \item For the NASA Parallel Benchmark suite, the maximum error is a negligent 3.46\%. This empirically establishes our completion time model's validity and applicability for complex, real-world applications as well as HPC environments.
\end{itemize}

With the maximum error of 6.81\%, our unified completion time model was able to predict the time for various CPU utilizations and frequency configurations across heterogeneous benchmarks and systems. Moreover, for the NASA Parallel Benchmark suite, our model predicted their completion time with a very high accuracy (maximum error of 3.46\%) and hence validating the applicability of our completion time model for complex applications and HPC environments. Thus, we proposed a completion time model to predict the performance of tasks operating inside the VMs using only 2 frequencies, a CPU fraction and 4 execution time values. We addressed the issue of predicting the running time for memory- and file-intensive benchmarks as well as HPC benchmarks and our model has a high prediction accuracy across heterogeneous virtualized systems. Our completion time model can also be used for Hadoop and HPC environments with no modifications. 

We would like to make a small note here. Performance of any task can be measured in one of two ways - given set of operations, how long does execution take OR given a fixed amount of time, how many operations are executed. We utilized benchmarks that complete execution to derive and validate our Completion Time model. On the other side are `unconstrained' applications (i.e., unconstrained w.r.t. time) which require throughput analysis rather than Completion Time modeling. For example, VM cloning \cite{SnowFlock} can be used to feed same request/input to all the clones operating at different CPU/frequency configurations and observing their performance. However, one potential way our model can be directly applied to such tasks is by identifying repeated subtasks or introducing breakpoints and observing their CT for different configurations. The following section presents the integration of our proposed models.

\section{Integration of Power and Completion Time Models}
\label{sec:integration}

The power and completion time models that we have empirically derived and validated in the previous sections can be integrated in multiple ways to provision VMs on physical machines (PMs). In this section, we describe two basic scenarios - (i) power-optimal provisioning and (ii) provisioning with a power budget.

\begin{figure}[!htbp]
\centering
\vspace{-0.3cm}
\includegraphics[scale=0.7]{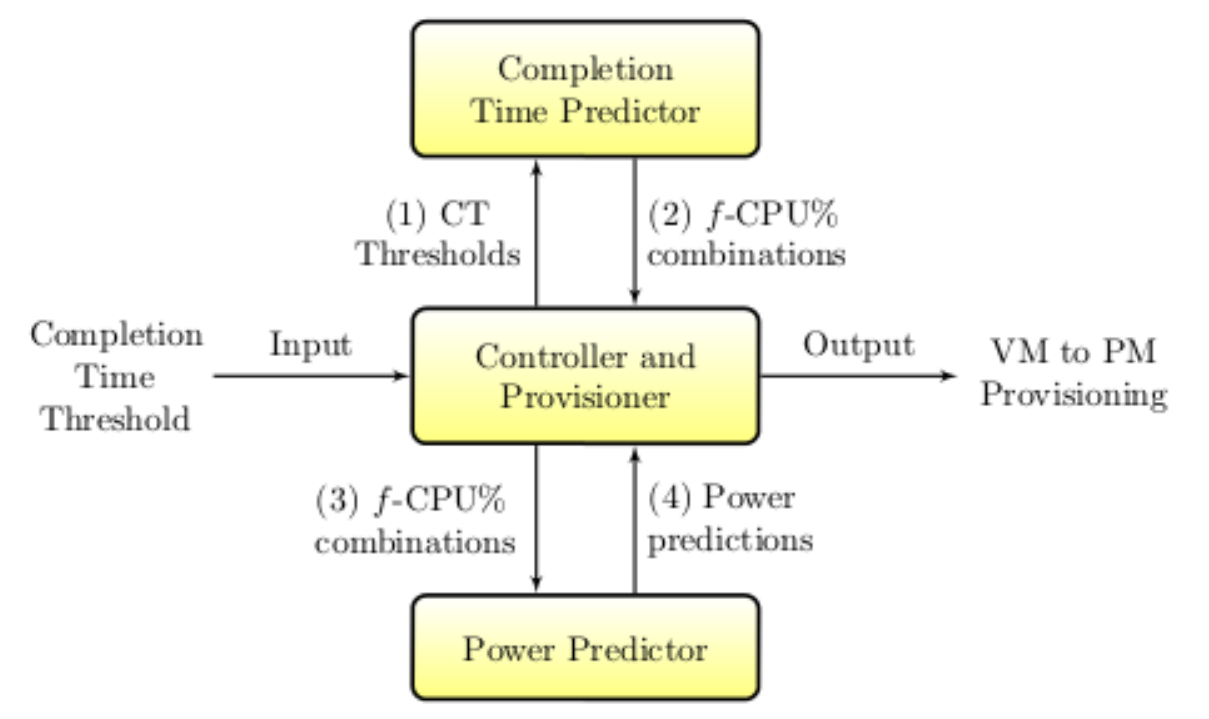}
\caption{Procedure for integrating our proposed models and power-optimally provisioning VMs}
\label{fig:provisioningFlowDiagram}
 \vspace{-0.5cm}
\end{figure}

\subsection{Scenario 1: Power-optimal provisioning}
\begin{figure}[!htbp]
\centering
\vspace{-0.3cm}
\includegraphics[scale=1]{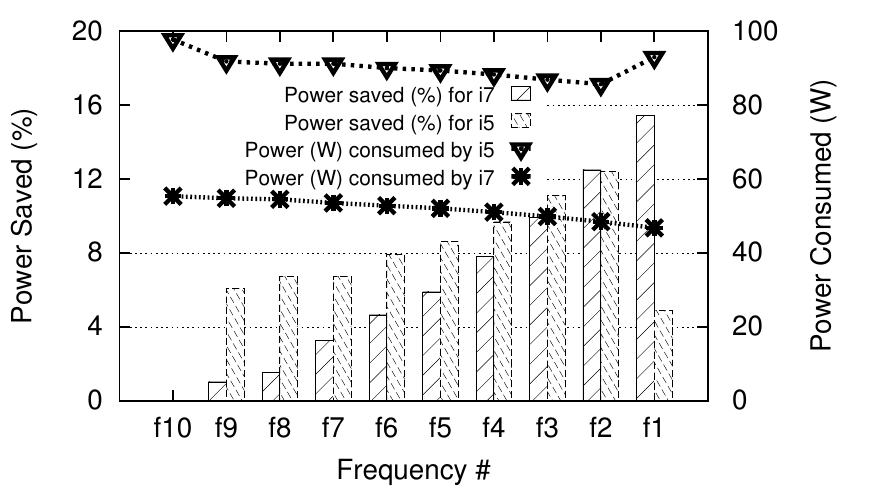}
\caption{Power saved and consumed by Intel i5 and i7 for a completion time threshold of 240s}
\label{fig:powersave}
 \vspace{-0.5cm}
\end{figure}

Consider a scenario where 4 VMs executing the same application - NumBench, have to be provisioned on a single PM in a power-optimal manner. The only input given is the completion time threshold ($CT_{Threshold}$) of 240 seconds. Assume that the completion time of the application is already characterized w.r.t. CPU\% and $f$ for Intel i5 and i7. Figure \ref{fig:provisioningFlowDiagram} shows the procedure for provisioning VMs such that the PM utilized the least amount of power. The $CT_{Threshold}$ is given as input to the Controller which forwards it to the Completion Time Predictor. This Predictor sends all feasible $f$-CPU\% combinations to the Controller. In order to provision in a power optimal manner, the Controller forwards the $f$-CPU\% combinations to the Power Predictor. The power characteristics of Intel i5 and i7 w.r.t. CPU\% and $f$ are already available with the Power Predictor. Therefore, the power utilized for the $f$-CPU\% combinations are calculated and sent to the Controller which chooses the least power consuming combination. Figure \ref{fig:powersave} shows the power saving achieved at each $f$ w.r.t. power consumed at $f_{10}$, the highest frequency. For a $CT_{Threshold}$ of 240 seconds, Intel i7 achieves 15.4\% i.e., 9W of power saving if all 4 VMs are provisioned at the lowest frequency of 1.6GHz and 71\% CPU. Whereas for Intel i5, the lowest $f$ of 1.2GHz yields only 5\% savings, but operating at 1.6GHz achieves 12.4\% i.e., 12W of saving. Hence, the Controller provisions the VMs on Intel i5 with 58\% CPU at 1.6GHz.

\subsection{Scenario 2: Provisioning with a power budget}
\begin{figure}[!htbp]
\centering
\includegraphics[scale=1]{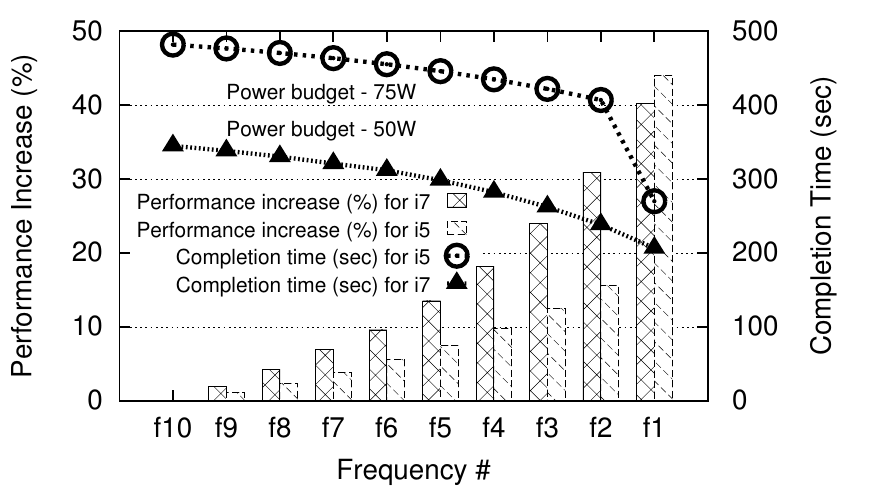}
 \vspace{-0.5cm}
\caption{Performance increased and achieved by Intel i5 and i7 for a power budget of 75W and 50W}
\label{fig:powerbudget}
\end{figure}

While one side of the coin is saving power with a performance constraint, improving performance for a power budget constraint is the other side. The objective for this scenario is to improve the performance i.e., reduce the completion time of the tasks executing inside VMs while maintaining the power utilized under a budget. The procedure for provisioning VMs for this scenario is similar to  the one shown in Figure \ref{fig:provisioningFlowDiagram} except the following changes.  A power budget is given as input which the Controller forwards to the Power Predictor. The feasible $f$-CPU\% combinations are sent to the Completion Time Predictor via the Controller. The Completion Time Predictor estimates the performance that will be achieved based on the characteristics of the application w.r.t. $f$ and CPU\% known apriori. Figure \ref{fig:powerbudget} shows the increase in the performance that is achieved by Intel i7 and i5 which have 50W and 75W respectively, as their power budgets. The completion time decreased i.e., the performance improved gradually with decrease in $f$ for Intel i7 and reached as much as 40\% (345s to 206s). The same trend was observed for Intel i5, where the performance improvement was 43.9\% (482s to 270s) with decrease in $f$.

In this section, we described two scenarios that utilized the integration of our power and completion time models. We explained the procedure for integrating our models and observed that as much as 15.4\% power can be saved and 43.9\% performance can be improved. 

\section{Conclusions}
\label{sec:conclusions}
In this paper, we proposed a power and a performance model with CPU utilization and frequency as parameters. To achieve that, we studied the power consumption trend of Intel i7, established that only the highest identified the parameters and empirically derived a power model with CPU allocation and frequency as input. We validated the power model by predicting power of 5 heterogeneous systems, with a maximum error of 7.4\%. We also derived a completion time model by combining existing models that predict using CPU allocation and frequency of the system. We validated the completion time model by predicting the execution time of 10 tasks on 4 machines with a maximum error of 6.8\%. We also described two scenarios where the integration of our proposed models achieved 15.4\% power saving and 43.9\% performance improvement while provisioning VMs. They can also be applied to HPC and Hadoop environments with no modifications.

For future work, we intend to extend the power model to dual or n-processor systems, encompass memory, file and network resources and propose a VM provisioning algorithm that is completion time-aware and power optimal across a cloud setup. The algorithm can also be extended to include dynamic SLA requirement and reallocation of server resources power optimally.

\bibliographystyle{abbrv}
\scriptsize{\bibliography{swethaICPE2015}}

\begin{thebibliography}{10}

\bibitem{ALM10}
\url{http://www.mes.co.in/images/resource/Downloads/ALM810Brochure.pdf}.

\bibitem{randmem}
\url{http://www.roylongbottom.org.uk/linux%20benchmarks.htm#anchor9}.

\bibitem{nasa}
\url{http://www.nas.nasa.gov/publications/npb.html}.

\bibitem{sysbench}
\url{http://sourceforge.net/projects/sysbench}.

\bibitem{cputest}
\url{http://tinyurl.com/ojtkf9m}.

\bibitem{kc}
\url{https://www.kernel.org/pub/linux/kernel/v3.x/linux-3.9.4.tar.xz}.

\bibitem{iozone}
\url{http://www.iozone.org}.

\bibitem{Amdahl1967}
G.~M. Amdahl.
\newblock Validity of the single processor approach to achieving large scale
  computing capabilities.
\newblock In {\em AFIPS}, 1967.

\bibitem{Bellosa}
F.~Bellosa.
\newblock The benefits of event-driven energy accounting in power-sensitive
  systems.
\newblock In {\em SIGOPS European Workshop}, 2000.

\bibitem{Benini2000}
L.~Benini, A.~Bogliolo, and G.~{De Micheli}.
\newblock {A survey of design techniques for system-level dynamic power
  management}.
\newblock {\em IEEE T VLSI SYST}, 2000.

\bibitem{Bertrana}
R.~Bertran, Y.~Becerra, D.~Carrera, V.~Beltran, M.~Gonzalez, X.~Martorell,
  J.~Torres, and E.~Ayguade.
\newblock Accurate energy accounting for shared virtualized environments using
  pmc-based power modeling techniques.
\newblock In {\em GRID}, 2010.

\bibitem{Mobius}
W.~Dargie, A.~Schill, and C.~Mobius.
\newblock Power consumption estimation models for processors, virtual machines,
  and servers.
\newblock {\em IEEE T PARALL DISTR}, 2014.

\bibitem{Dhiman2008}
G.~Dhiman, K.~K. Pusukuri, and T.~Rosing.
\newblock Analysis of dynamic voltage scaling for system level energy
  management.
\newblock In {\em HotPower}, 2008.

\bibitem{Fan02}
X.~Fan, C.~S. Ellis, and A.~R. Lebeck.
\newblock The synergy between power-aware memory systems and processor voltage
  scaling.
\newblock In {\em Power-Aware Computing Systems}, pages 164--179, 2002.

\bibitem{Fan2007}
X.~Fan, W.-D. Weber, and L.~A. Barroso.
\newblock Power provisioning for a warehouse-sized computer.
\newblock {\em SIGARCH Comput. Archit. News}, 2007.

\bibitem{Gurumurthi2002}
S.~Gurumurthi, A.~Sivasubramaniam, M.~J. Irwin, N.~Vijaykrishnan, M.~Kandemir,
  T.~Li, and L.~K. John.
\newblock Using complete machine simulation for software power estimation: The
  softwatt approach.
\newblock In {\em HPCA}, 2002.

\bibitem{Hsu2005}
C.-h. Hsu and W.-c. Feng.
\newblock A power-aware run-time system for high-performance computing.
\newblock In {\em SC}, 2005.

\bibitem{Isci2003}
C.~Isci and M.~Martonosi.
\newblock Runtime power monitoring in high-end processors: Methodology and
  empirical data.
\newblock In {\em MICRO}, 2003.

\bibitem{Kamga2011}
C.~M. Kamga, G.~S. Tran, and L.~Broto.
\newblock Power-aware scheduler for virtualized systems.
\newblock In {\em GCM}, 2011.

\bibitem{Kurd2009}
N.~Kurd, P.~Mosalikanti, M.~Neidengard, J.~Douglas, and R.~Kumar.
\newblock Next generation intel core micro-architecture (nehalem) clocking.
\newblock {\em IEEE J. of Solid-State Circuits}, 2009.

\bibitem{SnowFlock}
H.~A. Lagar-Cavilla, J.~A. Whitney, A.~M. Scannell, P.~Patchin, S.~M. Rumble,
  E.~de~Lara, M.~Brudno, and M.~Satyanarayanan.
\newblock Snowflock: Rapid virtual machine cloning for cloud computing.
\newblock In {\em EuroSys}, 2009.

\bibitem{Li2008}
K.~Li.
\newblock {Performance analysis of power-aware task scheduling algorithms on
  multiprocessor computers with dynamic voltage and speed}.
\newblock {\em IEEE T PARALL DISTR}, 2008.

\bibitem{Marinoni2007}
M.~Marinoni and G.~C. Buttazzo.
\newblock Elastic dvs management in processors with discrete voltage/frequency
  modes.
\newblock {\em IEEE T IND INFORM}, 2007.

\bibitem{Nathuji2007}
R.~Nathuji and K.~Schwan.
\newblock {Virtualpower : coordinated power management in virtualized
  enterprise systems}.
\newblock In {\em ICAC}, 2007.

\bibitem{Pedram2011}
M.~Pedram.
\newblock {Power and performance modeling in a virtualized server system}.
\newblock {\em ICPPW}, 2010.

\bibitem{Petrucci2011}
V.~Petrucci, E.~Carrera, O.~Loques, J.~C.~B. Leite, and D.~Mosse.
\newblock Optimized management of power and performance for virtualized
  heterogeneous server clusters.
\newblock In {\em CCGrid}, 2011.

\bibitem{Qureshi2009}
A.~Qureshi, R.~Weber, H.~Balakrishnan, J.~Guttag, and B.~Maggs.
\newblock Cutting the electric bill for internet-scale systems.
\newblock In {\em SIGCOMM}, 2009.

\bibitem{Rivoire2008}
S.~Rivoire, P.~Ranganathan, and C.~Kozyrakis.
\newblock A comparison of high-level full-system power models.
\newblock In {\em HotPower}, 2008.

\bibitem{Singh2009}
K.~Singh, M.~Bhadauria, and S.~A. McKee.
\newblock Real time power estimation and thread scheduling via performance
  counters.
\newblock {\em SIGARCH Comput. Archit. News}, 2009.

\bibitem{swetechreport}
S.~P.~T. Srinivasan and U.~Bellur.
\newblock A novel power model and completion time model for virtualized
  environments.
\newblock In {\em Technical Report, Dept. of CSE, IIT Bombay, TR-CSE-2014--58},
  2014.

\bibitem{Venkatachalam2006}
V.~Venkatachalam, M.~Franz, and C.~W. Probst.
\newblock A new way of estimating compute-boundedness and its application to
  dynamic voltage scaling.
\newblock {\em IJES}, 2007.

\bibitem{Wang2011}
X.~Wang and Y.~Wang.
\newblock {Coordinating Power Control and Performance Management for
  Virtualized Server Clusters}.
\newblock {\em IEEE T PARALL DISTR}, 2011.

\bibitem{Wen2010}
C.~Wen, J.~He, J.~Zhang, and X.~Long.
\newblock Pcfs: power credit based fair scheduler under dvfs for muliticore
  virtualization platform.
\newblock In {\em GreenCom CPSCom}, 2010.

\bibitem{Yadav}
M.~K. Yadav, M.~R. Casu, and M.~Zamboni.
\newblock {Dvfs based on voltage dithering and clock scheduling for gals
  systems}.
\newblock In {\em IEEE ASYNC}, 2012.

\bibitem{Zhao2011}
X.~Zhao and N.~Jamali.
\newblock {IGCC}.
\newblock {\em 2011 International Green Computing Conference and Workshops},
  2011.

\end{thebibliography}
 
\balancecolumns
\end{document}